\documentclass[sigplan,screen,natbib,nonacm]{acmart}

\usepackage[paperVersion]{dpupaper}
\AtBeginDocument{%
  \providecommand\BibTeX{{%
    \normalfont B\kern-0.5em{\scshape i\kern-0.25em b}\kern-0.8em\TeX}}}

\setcopyright{acmlicensed}
\acmPrice{15.00}
\acmDOI{10.1145/3385412.3385997}
\acmYear{2020}
\copyrightyear{2020}
\acmSubmissionID{pldi20main-p243-p}
\acmISBN{978-1-4503-7613-6/20/06}
\acmConference[PLDI '20]{Proceedings of the 41st ACM SIGPLAN International Conference on Programming Language Design and Implementation}{June 15--20, 2020}{London, UK}
\acmBooktitle{Proceedings of the 41st ACM SIGPLAN International Conference on Programming Language Design and Implementation (PLDI '20), June 15--20, 2020, London, UK}

\begin{document}

\title{\projName: Neural Type Hints}
\newcommand{\typeemb}{\vect{r}\!}

\author{Miltiadis Allamanis}
\email{miallama@micosoft.com}
\orcid{0002-5819-9900}
\affiliation{%
  \institution{Microsoft Research}
  \city{Cambridge}
  \country{United Kingdom}
}

\author{Earl T. Barr}
\email{e.barr@ucl.ac.uk}
\affiliation{%
  \institution{University College London}
  \city{London}
  \country{United Kingdom}}

\author{Soline Ducousso}
\authornote{Work done during an internship at UCL.}
\email{soline.ducousso@ensta-paristech.fr}
\affiliation{%
  \institution{\'{E}cole Nationale Sup\'{e}rieure de Techniques Avanc\'{e}es}
  \city{Paris}
  \country{France}
}

\author{Zheng Gao}
\email{z.gao.12@ucl.ac.uk}
\affiliation{%
 \institution{University College London}
 \city{London}
 \country{United Kingdom}}

\newcommand{\projName}{\mbox{\textsc{Typilus}}\xspace}
\newcommand{\projURL}{\mbox{\url{https://github.com/typilus/typilus}}\xspace}
\newcommand{\typespace}{\mbox{$\mathit{TypeSpace}$}\xspace}
\newcommand{\typemap}{\ensuremath\tau\hspace*{-0.4mm}\mathit{map}}

\renewcommand{\shortauthors}{Allamanis, \etal}

\begin{abstract}

Type inference over partial contexts in dynamically typed languages is
challenging. In this work, we present a graph neural network model that predicts
types by probabilistically reasoning over a program's structure, names, and
patterns. The network uses deep similarity learning to learn a \typespace --- a
continuous relaxation of the discrete space of types --- and how to embed the
type properties of a symbol (\ie identifier) into it. Importantly, our model can
employ one-shot learning to predict an open vocabulary of types, including rare
and user-defined ones. We realise our approach in \projName for Python that
combines the \typespace with an optional type checker. We show that \projName
accurately predicts types. \projName confidently predicts types for 70\% of all
annotatable symbols;  when it predicts a type, that type optionally type checks
95\% of the time. \projName can also find incorrect type annotations; two
important and popular open source libraries, \code{fairseq} and \code{allennlp},
accepted our pull requests that fixed the annotation errors \projName
discovered.

\end{abstract}



\maketitle

\section{Introduction}

Automatic reasoning over partial contexts is critical, since software
development necessarily involves moving code from an incomplete state ---
relative to the specification --- to a complete state. It facilitates scaling
analyses to large, industrial code bases. For example, it allows quickly
building a file-specific call graph when editing a single file instead of
requiring a project-level call graph. Traditional static analyses tackle partial
contexts via abstraction. They achieve soundness (or
soundiness~\citep{livshits2015defense}) at the cost of imprecision. As the
target language becomes more dynamic~\citep{richards2010analysis} (\eg
reflection or \lstinline|eval|), analysis precision tends to deteriorate.

Optional type systems~\cite{bracha2004pluggable} feature omittable type
annotations that have no runtime effects; they permit reasoning about the types
over a partial program that does not contain enough information to be fully
typed. In the limit, optional typing reaches traditional static typing when
all types are annotated or can be inferred. Recent years have witnessed
growing support of optional typing in dynamically typed languages. Notable
examples include \href{https://flow.org}{\code{flow}} for JavaScript and
\href{http://www.mypy-lang.org/}{\code{mypy}} for Python. This trend indicates
that the dynamic typing community increasingly acknowledges the
benefits of types, such as early bug detection~\cite{gao2017type}, better
performance, and accurate code completion and navigation.

Optional typing, however, comes with a cost. Developers who wish to
use it need to migrate an unannotated codebase to an (at least
partially) annotated one. When the codebase is large, manually adding type
annotations is a monumental effort. Type inference can automatically and
statically determine the most general type of a program expression, but it
offers little help here, because it cannot soundly deduce the types of
many expressions in a dynamic language, like JavaScript or Python.

In this work, we present \projName, a machine learning approach that suggests
type predictions to a developer.  Developers introduce a rich set of patterns into source
code, including the natural language elements, such as variable names, and
structural idioms in a program's control and data flows.  \projName exploits
these noisy and uncertain data, which traditional analysis ignores or
conservatively abstracts, to probabilistically suggest types.  Any suggestions
the developer accepts increase the number of typed terms in a program and
enhance the reasoning power of optional typing. Our goal is to help developers
move to fully typed programs and return to the safe haven of the guarantees that
traditional static typing affords.  Our work rests on the idea that verifying a
type suggestion helps a developer more quickly find the correct type than
coming up with the correct type from scratch.

Research has shown that machine learning methods can effectively predict
types~\citep{raychev2015predicting,hellendoorn2018deep}. This pioneering work,
however, treats this task as a classification problem and, therefore, have a
fixed set of categories, or types in our setting. In other words, they target a
\emph{closed} type vocabulary. However, the types in our corpus generally follow
a fat-tailed Zipfian distribution and 32\% of them are rare
(\autoref{sec:evaluation}). Type prediction for closed type vocabularies does
\emph{not} handle rare types and, thus, faces a performance ceiling.

\projName advances the state of the art by formulating probabilistic type
inference as a metric-based meta-learning problem, instead of a classification
problem. Meta-learning (or ``learning-to-learn'') allows trained machine
learning models to adapt to new settings with just a single example and no
additional training.  Metric-based meta-learning models learn to embed a (possibly) discrete
input into a real $D$-dimensional latent space preserving its properties.
Specifically, \projName learns to embed symbols (identifiers
like variable names, parameter names, function names \etc) capturing their
type properties.  We call these embeddings \emph{type embeddings}.
\projName strives to preserve the type equality relations among type embeddings
and establish, through training, a \typespace that keeps the type embeddings of
symbols that share a type close and those of different types apart.  \projName
predicts types for identifiers that lack type annotations because they cannot
be inferred and have not been provided, so it does not explicitly embed type annotations.

In contrast to classification-based methods, \projName can efficiently predict
types unseen during training.  Once we have a trained model, we can use it to
compute $\typeemb_s$, the type embedding of the symbol $s$ of a new type $\tau$
in the \typespace.  \projName maintains a type map from representative
embeddings to their type.  This type map implicitly defines ``typed'' regions in
the \typespace.  To allow \projName to predict $\tau$, we update this type map
with $\typeemb_s \mapsto \tau$. Now \projName can predict type $\tau$ for
symbols whose embeddings fall into a neighbourhood around
$\typeemb_s$ in its \typespace.  Thus, \projName can support an open type vocabulary,
\emph{without retraining}.  To bootstrap the process, we seed the \typespace and
the type map with the type embeddings of symbols with known types.  A type
checker removes false positives.  Our results show that \projName makes a
substantial advance, improving the state-of-the-art from 4.1\% to 22.4\%, when
predicting an exact match for rare types.

We realise \projName in a graph neural network (GNN).  Most existing neural
approaches treat programs as sequences of tokens, missing the opportunity to
model and learn from the complex dependencies among the tokens in a program. Our
experiment reveals that using a graph-based model, instead of a sequence-based
model, produces 7.6\% more exact matches for common types
(\autoref{tbl:typeresults}). We implement \projName to predict types for
variables, parameters, and function returns in a Python program.  For a given
symbol, two types are \emph{neutral} under an optional type system when
replacing one with the other does not yield a type error.  We show that, given
an appropriate confidence threshold that allows each model to predict types for
70\% of the symbols in our corpus, \projName' predictions achieve type
neutrality with the human-provided ones 95\% of the time, compared to 60\% for
the baseline model.

Human-written optional type annotations can fail to trigger a type error even
when they are incorrect.  \projName can find these errors.  In
\href{https://github.com/pytorch/fairseq}{\code{PyTorch/fairseq}}, a popular
sequence-to-sequence modelling toolkit, \projName predicted \code{int} when the
existing annotation was \code{float}; in
\href{https://github.com/allenai/allennlp}{\code{allenai/allennlp}}, a popular
natural language processing library, \projName predicted \code{Dict[str,ADT]}
when the existing annotation was \code{ADT}. We submitted two pull requests for
these errors; both were accepted (\autoref{sec:qual:eval}).
\projName and the evaluation artefacts can be found at \projURL.

\paragraph*{Contributions}
\begin{enumerate}
	\item We adapt a graph-based deep neural network to the type prediction problem
	by considering source code syntax and semantics;
	\item We use a novel training loss --- based on deep similarity learning --- to
	train a model that embeds the type properties of symbols into a \typespace and
	is adaptive:  without retraining, it can accurately predict types that were rare, or even
	unseen, during training;
	\item We realise our approach in \projName for Python and demonstrate its
	effectiveness in predicting types in an extensive evaluation.
\end{enumerate}

\section{Overview}

\begin{figure*}\centering
    \includegraphics[width=\textwidth]{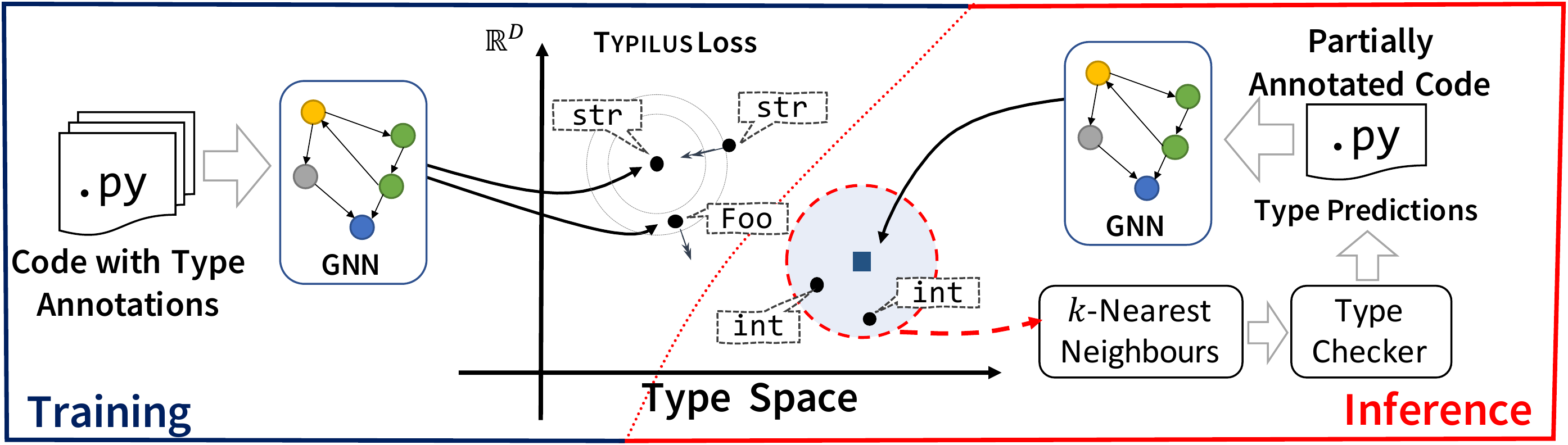}
    \caption{Overview of \projName. Training (\emph{left, blue}): A graph neural
    network (GNN) learns to map variables, parameters, and function returns to a type embedding
    in an $\mathbb{R}^D$ type space using deep similarity learning. Inference (\emph{right, red}): Using the type map, \projName
    accepts unannotated code, computes type embeddings with the trained GNN and finds the concrete $k$ nearest neighbours
    types as the candidate predictions. Finally, a type checker checks all
    predictions and filters incorrect ones.} \label{fig:arch}
\end{figure*}

In this work, we aim to predict types for symbols in an optionally typed
language.  This task takes two forms, open or closed, depending on whether one
aims to predict from a set of types that is finite and \emph{closed} or
unbounded and \emph{open}.  Early work for this task has targeted a closed type
vocabulary; DeepTyper considers 11.8k types found in a large TypeScript corpus
where JSNice considers a smaller
set~\citep{raychev2015predicting,hellendoorn2018deep}. However, most real-life
code introduces new, domain-specific types. In our data
(\autoref{sec:evaluation}), only 158 types (out of 49k) appear more than 100
times, following a Zipfian distribution, similar to other code
artefacts~\citep{allamanis2013mining}. Nevertheless, given the fat-tailed
distribution of types, about 32\% of the type annotations are rare in our corpus. Thus,
predicting from a closed type vocabulary faces a performance ceiling.

We present \projName, a method targeting an open type vocabulary that can
predict types unseen during training. \autoref{sec:evaluation} shows that many
of these predictions are useful because an type checker can efficiently verify
them; when they are not type checkable, we hope they at least speed a
developer's selection of the correct type. \autoref{fig:arch} depicts the
high-level architecture of \projName.

\emph{\textbf{Learning a Type Space}} (\autoref{fig:arch}, \emph{left,
blue})\quad Central to \projName is its \typespace, a continuous projection of
the type context and properties of code elements into a real multidimensional space.  We do
\emph{not} explicitly design this space, but instead learn it from data.  To
achieve this, we train a neural network $e(\cdot)$ that takes a code snippet $S$
and learns to map the variables, parameters and functions of $S$ into the
\typespace. For each symbol $s \in S$, $e(S)[s] = \typeemb_s \in \mathbb{R}^D$.
\projName uses deep similarity learning,  which needs sets of positive and
negative examples for training.  To define these sets, we leverage existing type
annotations in Python programs: like \code{str} and \code{Foo} in
\autoref{fig:arch}.  To capture both semantic and lexical properties of code,
\projName employs a graph neural network (GNN). The GNN learns and
integrates information from multiple sources, including identifiers, syntactic
constraints, syntactic patterns, and semantic properties like control
and data flow.

\emph{\textbf{Predicting Types}} (\autoref{fig:arch}, \emph{right, red})\quad
\projName's $e$ alone cannot directly predict a type, since it maps symbols to
their type embeddings in $\mathbb{R}^D$.  Instead, its output, the \typespace,
acts as an intermediate representation between a program's symbols and their
concrete types. To make type predictions, \projName builds $\typemap$ to
map a symbol's type embedding to its type. This implicitly maps each type to a set of 
type embeddings. First, given a corpus of
code --- not necessarily the training corpus --- we map all known type
annotations $\tau_i$ into points in the \typespace. In \autoref{fig:arch},
\code{int} is an example $\tau_i$. Given the $\typemap$ and a query symbol $s_q$
(the \emph{black square} in \autoref{fig:arch}), whose type properties $e$
embeds at $\typeemb_{s_q}$ (\ie $e(S)[s_q] = \typeemb_{s_q}$), \projName returns
a probability distribution over candidate type predictions in the neighbour
around $\typeemb_{s_q}$ in the \typespace.  \projName uses $k$ nearest
neighbours to define this neighbourhood.  Finally, a type checker checks the
highest-probability type predictions and, if no type errors are found, \projName
suggests them to the developer.

\paragraph*{Key Aspects}
The key aspects of \projName are
\begin{itemize}
    \item A graph-based deep neural network that learns from the rich set of
    patterns and can predict types without the need to have a fully resolved
    type environment.
    \item The ability to embed the type properties of any symbol, including symbols
    whose types were unseen during training, thereby tackling the problem of
    type prediction for an open vocabulary.
    \item A type checking module that filters false positive type predictions,
    returning only type-correct predictions.
\end{itemize}

\section{Background}
\label{sec:background}

Machine learning has been recently used for predicting type
annotations in dynamic languages. Although traditional type inference is
a well-studied problem, commonly it cannot handle
uncertainty as soundness is an important requirement. In contrast,
probabilistic machine learning-based type inference employs
probabilistic reasoning to assign types to a program. For
example, a free (untyped) variable named \code{counter} will never
be assigned a type by traditional type inference,
whereas a probabilistic method may infer that given the
name of the variable an \code{int} type is highly likely.

\citet{raychev2015predicting} first introduced probabilistic type inference
using statistical learning method for JavaScript code in their tool JSNice.
JSNice converts a JavaScript file into a graph representation encoding some
relationships that are relevant to type inference, such as relations among
expressions and aliasing. Then a conditional random field (CRF) learns to
predict the type of a variable. Thanks to the CRF, JSNice captures constraints
that can be statically inferred.  Like all good pioneering work, JSNice poses
new research challenges.  JSNice targets JavaScript and a closed vocabulary of
types, so predicting types for other languages and an open type vocabulary are
two key challenges it poses. \projName tackles these challenges.  \projName
targets Python.  To handle an open vocabulary, \projName exploits subtokens,
while JSNice considers tokens atomic. \projName replaces the CRF with a graph
neural network and constructs a \typespace.

Later, \citet{hellendoorn2018deep} employed deep learning to solve this problem.
The authors presented DeepTyper, a deep learning model that represents the
source code as a sequence of tokens and uses a sequence-level model (a biLSTM)
to predict the types within the code. In contrast to
\citet{raychev2015predicting}, this model does not explicitly extract or
represent the relationships among variables, but tries to learn these
relationships directly from the token sequence. DeepTyper performs well and can
predict the types with good precision. Interestingly,
\citet{hellendoorn2018deep} showed that combining DeepTyper with JSNice improves
the overall performance, which suggests that the token-level patterns and the
extracted relationships of \citet{raychev2015predicting} are complementary. Our
work incorporates these two ideas in a Graph Neural Network. Similar to
\citet{raychev2015predicting}, DeepTyper cannot predict rare or previously
unseen types and its implementation does not subtokenise identifiers.

\citet{malik2019nl2type} learn to predict types for JavaScript code just from
the documentation comments. They find that (good) documentation comments (\eg
docstrings in Python) contain valuable information about types. Again, this is
an ``unconventional'' channel of information that traditional type inference
methods would discard, but a probabilistic machine learning-based method can
exploit.  Like other prior work, \citet{malik2019nl2type}'s approach suffers
from the rare type problem.  It is orthogonal to \projName, since \projName
could exploit such sources of information in the future.

Within the broader area --- not directly relevant
to this work --- research has addressed other problems that combine machine
learning and types. \citet{dash2018refinym} showed that, by exploiting the natural
language within the names of variables and functions, along
with static interprocedural data flow information, existing typed
variables (such as \code{string}s) can be nominally refined.
For example, they automatically refine string variables into
some that represent passwords, some that represents file system
paths, \etc Similarly,
\citet{kate2018phys} use probabilistic inference to predict
physical units in scientific code. While they use probabilistic
reasoning to perform this task, no learning is employed.

\section{The Deep Learning Model}

We build our type space model using deep learning, a versatile
family of learnable function approximator methods that is widely
used for pattern recognition in computer vision and natural
language processing~\citep{goodfellow2016deep}.
Deep learning models have three basic ingredients:
(a) a deep learning architecture tailored for learning task data (\autoref{subsec:gnns});
(b) a way to transform that data into a format that the
neural network architecture can consume (\autoref{sec:implementation}); and
(c) an objective function for training the neural network (\autoref{subsec:type embeddings}).
Here, we detail a deep learning model that solves the type prediction task for an
open type vocabulary, starting from (c) --- our objective function.
By appropriately selecting our objective function we
are able to learn the type space, which is the
central novelty of \projName.

\subsection{Learning a Type Space}
\label{subsec:type embeddings}

Commonly, neural networks represent their elements as ``distributed vector
representations'', which distribute the ``meaning'' across vector components.
A neural network may compute these vectors.  For the purpose of explanation,
assume a neural network $e(\cdot)$, parameterised by some learnable parameters
$\vect{\theta}$, accepts as input some representation of a code snippet $S$ and
returns a vector representation $\typeemb_s \in \mathbb{R}^D$ for each symbol $s
\in S$. We call this vector representation a \emph{type embedding}; it captures
the relevant type properties of a symbol in $S$. Below, we treat $e$ as a map
and write $e(S)[s] = \typeemb_s$ to denote $s$'s type embedding under $e$ in
$S$.  The type prediction problem is then to use type embeddings to predict the
type of a symbol. In \autoref{subsec:gnns}, we realise $e(\cdot)$ as a GNN.

A common choice is to use type embeddings 
for classification. For this purpose, we need a
finite set of well-known types $\mathcal{T}=\{\tau_i\}$. For each
of these types, we must learn a ``prototype'' vector representation $\tilde{\typeemb}_{\tau_i}$
and a scalar bias term $b_{\tau_i}$. Then,
given a ground truth type $\tau$ for a symbol $s$
with computed type embedding $\typeemb_s=e(s)$ we seek
to maximise the probability $P(s:\tau)$, \ie
minimise the classification loss
\begin{align} \label{eq:softmax classification}
    \mathcal{L}_{\textsc{Class}}\left(\typeemb_s, \tau\right) =  \underbrace{-\log \frac{\exp\left(\typeemb_s \tilde{\typeemb}_{\tau}^T+ b_{\tau}\right)}{\sum_{\tau_j \in \mathcal{T}} \exp{\left(\typeemb_s \tilde{\typeemb}_{\tau_j}^T+ b_{\tau_j}\right)}}}_{-\log P(s:\tau)}.
\end{align}
As the reader
can observe, \autoref{eq:softmax classification} partitions the
space of the type embeddings into the set of
well-known types via the prototype vector representations $\tilde{\typeemb}_{\tau_i}$.
This limits the model
to a closed vocabulary setting, where it
can predict only over a fixed
set of types $\mathcal{T}$. Treating type suggestion as
classification misses
the opportunity for the model to learn about
types that are rare or previously unseen, \eg in a new project.
In this work, we focus on ``open vocabulary'' methods, \ie methods
that allow us to arbitrarily expand the set of candidate type that
we can predict at test-time. We achieve this through similarity
learning, detailed next.

\paragraph*{Deep Similarity Learning}
\label{subsec:deep sim learning}
We treat the creation of a type space
as a weakly-supervised similarity learning problem~\citep{chopra2005learning,hadsell2006dimensionality}.  This process projects the discrete
(and possibly infinite) set of types into a multidimensional
real space. Unlike classification, this process does \emph{not}
explicitly partition the space in a small set.
Instead, $e(\cdot)$ learns to represent an open vocabulary of types, since
it suffices to map any new, previously unseen, symbol (and its type)
into some point in the real space. Predicting types for these symbols becomes
a similarity computation between the queried symbol's type embedding and nearby
embeddings in the type space; it does \emph{not} reduce to determining into which partition
a symbol maps, as in classification.

\newcommand{\norm}[1]{\left| \left| #1 \right| \right|}
To achieve this, we adapt a deep similarity learning method called \emph{triplet loss}~\citep{cheng2016person}.
The standard formulation of triplet loss accepts a type embedding $\typeemb_s$ of
a symbol $s$, an embedding $\typeemb_{s^+}$ of a symbol $s^+$ of the same type
as $s$ and an embedding $\typeemb_{s^-}$ of a symbol $s^-$ of a different
type than $s$. Then, given a positive scalar margin $m$, the triplet loss is
\begin{align} \label{eq:simpleTriplet}
    \mathcal{L}_{\textsc{Triplet}}(\typeemb_s, \typeemb_{s^-}, \typeemb_{s^+}) = h \left(\norm{\typeemb_{s} - \typeemb_{s^-}} - \norm{\typeemb_{s} - \typeemb_{s^+}}, m\right),
\end{align}
where $h(x, m) = max(x + m, 0)$ is the hinge loss. This objective
aims to make $s_i$'s embedding closer to
the embedding of the ``similar''
example $s_i^+$ than to the embedding of $s_i^-$, up to the margin $m$.
In this work, we use the $L_1$ (Manhattan) distance, but other distances can be used.
Learning over a similarity loss can be thought of as loosely analogous to a physics simulation
where each point exerts an attraction force on similar points (proportional
to the distance) and a repelling force (inversely proportional to the distance) to dissimilar points.
Triplet loss has been used for many applications such as the computer vision problem of
recognising if two hand-written signatures were signed by the same
person and for face recognition~\citep{cheng2016person}.
As \autoref{eq:simpleTriplet} shows, triplet loss merely requires that we define the pairwise
(dis)similarity relationship among any two samples, but does not require any concrete labels.

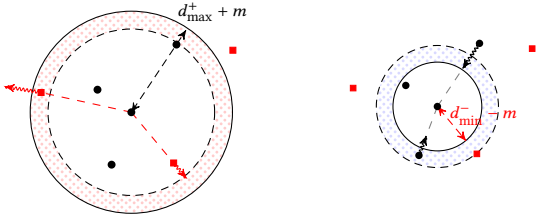
\begin{figure}[t]
    \begin{minipage}{0.48\columnwidth}
\centering
\begin{tikzpicture}[scale=0.75]
    \tikzset{
        >=stealth',
        force/.style={red,decorate,decoration={snake, segment length=.5mm, amplitude=0.2mm}}
    }
    \coordinate (ground) at (0,0);
    \fill[pattern=crosshatch dots, pattern color=red!20,even odd rule] (ground) circle (42pt) (ground) circle (51pt);
    \draw[black,densely dashed] (ground) circle (42pt);
    \draw[black] (ground) circle (51pt);

    \coordinate (pos1) at (0.8,1.2);
    \coordinate (pos2) at (-0.6,0.4);
    \coordinate (pos3) at (-0.35,-0.93);

    \coordinate (neg1) at (1.8,1.1);
    \coordinate (neg2) at (0.75,-0.9);
    \coordinate (neg3) at (-1.6,0.35);

    \draw[black, <->, densely dashed] (ground) -- (0.97,1.457) node[above, xshift=10pt] {\tiny$d^+_{\text{max}}+m$};
    \draw[red, dashed] (ground) -- (neg2); 
    \draw[->,force] (neg2) -- (0.975,-1.17);
    \draw[red, dashed] (ground) -- (neg3); 
    \draw[->,force] (neg3) -- (-2.24,0.455);

    \fill[black] (ground) circle [radius=2pt];
    \fill[black] (pos1) circle [radius=2pt];
    \fill[black] (pos2) circle [radius=2pt];
    \fill[black] (pos3) circle [radius=2pt];
   
    \filldraw[red] ([xshift=-1.5pt,yshift=-1.5pt]neg1) rectangle ++(3pt,3pt);
    \filldraw[red] ([xshift=-1.5pt,yshift=-1.5pt]neg2) rectangle ++(3pt,3pt);
    \filldraw[red] ([xshift=-1.5pt,yshift=-1.5pt]neg3) rectangle ++(3pt,3pt);
\end{tikzpicture}
\end{minipage}
\begin{minipage}{0.48\columnwidth}
\centering
\begin{tikzpicture}[scale=0.7]
    \tikzset{
        >=stealth',
        force/.style={black,decorate,decoration={snake, segment length=.5mm, amplitude=0.2mm}}
    }
    \coordinate (ground) at (0,0);
    \coordinate (pos1) at (0.8,1.2);
    \coordinate (pos2) at (-0.6,0.4);
    \coordinate (pos3) at (-0.35,-0.93);

    \coordinate (neg1) at (1.8,1.1);
    \coordinate (neg2) at (0.75,-0.9);
    \coordinate (neg3) at (-1.6,0.35);

    \fill[pattern=crosshatch dots, pattern color=blue!15,even odd rule] (ground) circle (24pt) (ground) circle (33pt);
    \draw[black,densely dashed] (ground) circle (33pt);
    \draw[black] (ground) circle (24pt);
    
    \draw[red, <->, densely dashed] (ground) -- (0.545,-0.654) node[midway,xshift=12,yshift=3] {\tiny$d^-_{\text{min}}-m$};
    \draw[black!50, dashed] (ground) -- (pos1); 
    \draw[->,force] (pos1) -- (0.48,0.72);
    \draw[black!50, dashed] (ground) -- (pos3); 
    \draw[->,force] (pos3) -- (-0.21,-0.558);

    \fill[black] (ground) circle [radius=2pt];
    \fill[black] (pos1) circle [radius=2pt];
    \fill[black] (pos2) circle [radius=2pt];
    \fill[black] (pos3) circle [radius=2pt];
   
    \filldraw[red] ([xshift=-1.5pt,yshift=-1.5pt]neg1) rectangle ++(3pt,3pt);
    \filldraw[red] ([xshift=-1.5pt,yshift=-1.5pt]neg2) rectangle ++(3pt,3pt);
    \filldraw[red] ([xshift=-1.5pt,yshift=-1.5pt]neg3) rectangle ++(3pt,3pt);
\end{tikzpicture}
\end{minipage}
    \caption{Graphic depiction of the two terms of the similarity objective in \autoref{eq:modifiedTriplet}.
             \emph{Left}: all dissimilar points (red squares), \ie $P_-$
             within distance $d^+_\text{max}+m$ of the query point are pushed away. \emph{Right}:
             all similar points (black circles) that are further than $d^-_\text{min}-m$
             from the query point, \ie $P_+$, are pulled towards it. The margin distance $m$ is shaded.
             }\label{fig:tripletloss}
\end{figure}

\paragraph*{\projName Loss}
\projName adapts triplet loss (\autoref{eq:simpleTriplet})
for the neural type representations to facilitate
learning and combines it with a classification-like loss.
  \autoref{fig:tripletloss} depicts the similarity loss conceptually.
\projName' similarity loss considers more
than three samples each time. Given a symbol $s$ and a set of similar $S^+_i(s)$
and a set of dissimilar $S^-_i(s)$ symbols drawn randomly from the dataset, let
\begin{align*}
	d^+_\text{max}(s) = \max_{ s^+_i \in S^{+}_i(s) }  \norm{\typeemb_s - \typeemb_{s^+_i}},\hspace{1em}
	d^-_\text{min}(s) = \min_{ s^-_i \in S^{-}_i(s) } \norm{\typeemb_s - \typeemb_{s^-_i}},
\end{align*}
\ie the maximum (resp. minimum) distance among identically (resp. differently) typed symbols.
Then, let
\begin{align*}
P_+(s) &= \left\{x^+_i: \norm{\typeemb_{s^+_i} - \typeemb_s }> d^-_\text{min}(s) - m\right\}, \\
P_-(s) &= \left\{x^-_i: \norm{\typeemb_{s^-_i} - \typeemb_s }< d^+_\text{max}(s) + m\right\}
\end{align*}
\ie
the sets of same and differently typed symbols that are within a margin of $d^+_\text{max}$
and $d^-_\text{min}.$ Then, we define the similarity loss over the type space as
\begin{align} \label{eq:modifiedTriplet}
    \mathcal{L}_{\textsc{Space}}(s) = \sum_{s^+_i \in P_+(s)} \frac{\norm{\typeemb_{s^+_i} - \typeemb_{s}}}{\left|P_+(s)\right|} -
                  \sum_{s^-_i \in P_-(s)} \frac{\norm{\typeemb_{s^-_i} - \typeemb_{s}}}{|P_-(s)|},
\end{align}
which generalises \autoref{eq:simpleTriplet} to multiple symbols. Because we can
efficiently batch the above equation in a GPU, it provides an alternative to
\autoref{eq:simpleTriplet} that converges faster, by reducing the sparsity of
the objective. In all our experiments, we set $S^+(s)$ (resp. $S^-(s)$) to the
set of symbols in the minibatch that have the same (resp. different) type as
$s$.

$\mathcal{L}_{\textsc{Class}}$ and $\mathcal{L}_{\textsc{Space}}$ have
different advantages and disadvantages.  $\mathcal{L}_{\textsc{Class}}$'s
prototype embeddings $\tilde{\typeemb}_{\tau_i}$ provide a central point of
reference during learning, but cannot be learnt for rare or previously unseen
types.  $\mathcal{L}_{\textsc{Space}}$, in contrast, explicitly handles pairwise
relationships even among rare types albeit at the cost of potentially reducing 
accuracy across all types: due to the sparsity of the pairwise
relationships, the \typespace may map the same type to different regions.
To construct more robust type embeddings, \projName combines both
losses in its learning objective:
\begin{align} \label{eq:projLoss}
    \mathcal{L}_{\projName}(s, \tau) = \mathcal{L}_{\textsc{Space}}(s) +
        \lambda \mathcal{L}_{\textsc{Class}}\left(W \typeemb_s, \textsc{Er}\left(\tau\right)\right),
\end{align}
where $\lambda=1$ in all our experiments, $W \typeemb_s$ is the type
embedding of $s$ in a linear projection of \projName's \typespace, and
$\textsc{Er}(\cdot)$ erases all type parameters.

\projName employs type erasure in \autoref{eq:projLoss} on parametric types to
combat sparsity.  When querying $\textsc{Er}(\tau)$, applying
$\mathcal{L}_{\textsc{Class}}$ directly to $\typeemb_s$ would risk collapsing
generic types into their base parametric type, mapping them in same location.
$W$ counteracts this tendency; it is a learned matrix (linear layer) that can be
thought as a function that projects the \typespace into a new latent space with
no type parameters. This parameter-erased space provides coarse information
about type relations among parametric types by imposing a linear relationship
between their type embeddings and their base parametric type; $W$ learns, for
instance, a linear relationship from \code{List[int]} and \code{List[str]} to
\code{List}. 

At inference time, \projName discards the prototype embeddings
$\tilde{\typeemb}_{\tau_i}$ and $W$.  \projName uses these
components of $\mathcal{L}_{\textsc{Class}}$ to learn the \typespace, which
retains them implicitly in its structure.  \autoref{tbl:typeresults} in
\autoref{sec:evaluation} presents and compares the performance of all the loss
functions discussed here.

\subsection{Adaptive Type Prediction}

Once trained, $e(\cdot)$ has implicitly learned a type space. However, the
type space does not explicitly contain types, so, for a set of symbols whose
types we know, we construct a map from their type embeddings to their types.
Formally, for every symbol $s$ with known type $\tau$, we use the trained
$e(\cdot)$ and add type markers to the type space, creating a map
$\typemap[e(S)[s]] \mapsto \tau$.

To predict a type for a query symbol $s_q$ in the code snippet $S$, \projName
computes $s_q$'s type embedding $\typeemb_{s_q}=e(S)[s_q]$, then finds the $k$
nearest neighbours ($k$NN) over $\typemap$'s keys, which are type
embeddings.  Given the $k$ nearest neighbour markers $\tau_i$
with a distance of $d_i$ from the query type embedding $\typeemb_{s_q}$, 
the probability of $s_q$ having a type $\tau'$ is
\begin{align} \label{eq:knn prob} P(s_q: \tau') = \frac{1}{Z} \sum_i
\mathbb{I}\left(\tau_i = \tau'\right) d_i^{-p} \end{align}
where $\mathbb{I}$ is the indicator function and $Z$ a normalising constant.
$p^{-1}$ acts as a temperature with $p \rightarrow 0$ yielding a uniform
distribution over the $k$ nearest neighbours and $p \rightarrow \infty$ yielding
the $k=1$ nearest neighbour algorithm.

Though $e(\cdot)$ is fixed, $\typemap$ is adaptive: it can accept bindings from
type embeddings to actual types.  Notably, it can accept bindings for previously
unseen symbols, since $e(\cdot)$ can compute an embedding for a previously
unseen $s$.  \projName' use of $\typemap$, therefore, allows it to adapt and
learn to predict new types without retraining $e$.  A developer or a type inference
engine can add them to $\typemap$ before test time.  \projName handles an
open type vocabulary thanks to the adaptability that $\typemap$ affords.

\paragraph*{Practical Concerns}
The $k$NN algorithm is costly if na{\"i}vely implemented. Thankfully,
there is a rich literature for spatial indexes that reduce the
time complexity of the $k$NN from linear to constant.
We create a spatial index of the type space and the relevant markers.
\projName then efficiently performs nearest-neighbour queries by
employing the spatial index. In this work, we employ
\href{https://github.com/spotify/annoy}{Annoy}~\citep{annoy} with $L_1$ distance.

\subsection{Graph Neural Network Architectures}
\label{subsec:gnns}

So far, we have assumed that some neural network $e(\cdot)$ can
compute a type embedding  $\typeemb_{s},$
but we have not defined this network yet.  A large set of
options is available; here, we focus on graph-based models.
In \autoref{sec:evaluation}, we consider also token- and AST-level models
as baselines.

Graph Neural Networks~\citep{li2015gated,kipf2016semi} (GNN) are a form of
neural network that operates over graph structures. The goal of a GNN is to
recognise patterns in graph data, based both on the data within the nodes and
the inter-connectivity. There are many GNN variants. Here, we describe the broad
category of message-passing neural networks~\citep{gilmer2017neural}. We then
discuss the specific design of the GNN that we employ in this work. Note that
GNNs should \emph{not} be confused with Bayesian networks or factor graphs,
which are methods commonly used for representing probability distributions and
energy functions.

Let a graph $G = (N, \mathcal{E})$ where $N=\{n_i\}$ is a set of nodes and
$\mathcal{E}$ is a set of directed edges of the form $n_i
\overset{k}{\rightarrow} n_j$ where $k$ is the edge label. The nodes and edges
of the graph are an input to a GNN. In neural message-passing GNNs, each node
$n_i$ is endowed with vector representation $\vect{h}_{n_i}^t$ indexed over a
timestep $t$. All node states are updated as
\newcommand{\aggOp}{\ensuremath{\bigoplus}}
\begin{align} \label{eq:gnn message passing}
    \vect{h}_{n_i}^{t+1} = f_t\left(\vect{h}_{n_i}^t, \aggOp_{\forall n_j: n_i \overset{k}{\rightarrow} n_j}\left(m^t(\vect{h}_{n_i}^t, k, \vect{h}_{n_j}^t)\right)\right),
\end{align}
where $m^t(\cdot)$ is a function that computes a ``message'' (commonly a vector) based on the edge label $k$,
$\aggOp$ is a commutative (message) aggregation operator
that summarises all the messages
that $n_i$ receives from its direct neighbours, and $f_t$ is an update function
that updates the state of node $n_i$.  We use parallel edges with different
labels between two nodes that share multiple properties.
The $f_t$, $\aggOp$ and $q^t$ functions contain
all trainable graph neural network parameters.
Multiple options are possible for $f_t$, $\aggOp$ and $q^t$s.
The initial state of each node $h_{n_i}^0$ set from node-level
information.  \autoref{eq:gnn message passing} updates all node states
$T$ times recursively. At
the end of this process, each $h_{n_i}^T$ represents information
about the node and how it ``belongs'' within the context
of the graph.

In this work, we use the gated graph neural network (GGNN) variant~\citep{li2015gated}
that has been widely used in machine learning models of source code. Although other
GNN architectures have been tested, we do not test them here. GGNNs
use a single GRU cell~\citep{cho2014properties} for all $f_t$, \ie $f_t=\textsc{Gru}(\cdot, \cdot)$,
$\aggOp$ is implemented as a summation operator and
$m^t(\vect{h}_{n_i}^t, k, \vect{h}_{n_j}^t)=E_k  \vect{h}_{n_j}^t$,
where $E_k$ is a learned matrix \ie $m^t$ is a linear layer that does not depend on $t$ or $\vect{h}_{n_i}^t$.

Overall, we
follow the architecture and hyperparameters used in
\citet{allamanis2018learning,brockschmidt2018generative}. \eg set $T=8$. In
contrast to previous work, we use max pooling (elementwise maximum) as the message aggregation
operator $\aggOp$. In early experiments, we found that max pooling performs somewhat
better and is conceptually better fitted to our setting since max pooling
can be seen as a meet-like operator over a lattice defined in $\mathbb{R}^N$.
Similar to previous work, $\vect{h}_{n_i}^0$ is defined as the average of the
(learned) subtoken representations of each node, \ie
\begin{align}
\vect{h}_{n_i}^0 = \frac{1}{|\textsc{SubTok}(n_i)|}\sum_{s \in \textsc{SubTok}(n_i)} \vect{e}_s,
\end{align}
where $\textsc{SubTok}(\cdot)$ deterministically splits the identifier information
of $n_i$ into subtokens on CamelCase and under\_scores and $\vect{e}_s$ is
an embedding of the subtoken $s$, which
is learned along with the rest of the model parameters.

\section{\projName: A Python Implementation}
\label{sec:implementation}

We implement \projName for Python, a popular dynamically typed
language. We first introduce Python's type hints (\ie annotations), then
describe how we convert Python code into a graph format.
Python was originally designed without a type annotation syntax. But starting from
version 3.5, it has gradually introduced language features that support
(optional) type annotations
(\href{https://www.python.org/dev/peps/pep-0484/}{PEP 484} and
\href{https://www.python.org/dev/peps/pep-0526/}{PEP 526}). Developers can now
optionally annotate variable assignments, function arguments, and returns. These
type annotations are not checked by the language (\ie Python remains dynamically
typed), but by a standalone type checker. The built-in
\href{https://docs.python.org/3/library/typing.html}{\code{typing}} module
provides some basic types, such as \code{List}, and utilities, like type
aliasing. We refer the reader to the
\href{https://docs.python.org/3/library/typing.html}{documentation}~\citep{pythonTyping}
for more information.

\subsection{Python Files to Graphs}
\label{subsec::graph:construction}

Representing code in graphs involves multiple design decisions. Inspired by
previous work
\cite{allamanis2018learning,alon2018code2vec,brockschmidt2018generative,cvitkovic2018open,raychev2015predicting},
our graph encodes the tokens, syntax tree, data flow, and symbol table of each
Python program and can be thought as a form of feature extraction. As such, the
graph construction is neither unique nor ``optimal''; it encapsulates design
decisions and trade-offs. We adopt this construction since it has been
successfully used in existing machine learning-based work. Traditionally, formal
methods discard many elements of our graph.  However, these elements are a
valuable source of information, containing rich patterns that a machine learning
model can learn to detect and employ when making predictions, as our results
demonstrate.

\begin{table*}
    \begin{tabular}{lp{12cm}l} \toprule
    Edge & This edge connects ...\\ \midrule
    \code{NEXT\_TOKEN} &  two consecutive token nodes. & \citep{allamanis2018learning,hellendoorn2018deep} \\
    \code{CHILD} & syntax nodes to their children nodes and tokens. & \citep{allamanis2018learning,alon2018code2vec,raychev2015predicting}\\
    \code{NEXT\_MAY\_USE}  & each token that is bound to a variable to all potential next uses of the variable. & \citep{allamanis2018learning}\\
    \code{NEXT\_LEXICAL\_USE} & each token that is bound to a variable to its next lexical use. &  \citep{allamanis2018learning}\\
    \code{ASSIGNED\_FROM} & the right hand side of an assignment expression to its left hand-side. & \citep{allamanis2018learning}\\
    \code{RETURNS\_TO} & all \code{return}/ \code{yield} statements to the function declaration node where control returns. & \citep{allamanis2018learning}\\
    \code{OCCURRENCE\_OF} & all token and syntax nodes that bind to a symbol to the respective symbol node. & \citep{cvitkovic2018open,gilmer2017neural}\\
    \code{SUBTOKEN\_OF} & each identifier token node to the vocabulary nodes of its subtokens. & \citep{cvitkovic2018open}\\
    \bottomrule
    \end{tabular}
    \caption{Description of edge labels used in our graph representation of Python. \autoref{fig:sampleGraph} shows a sample graph.}\label{tbl:edges}
\end{table*}

\begin{figure}
    \centering
    \includegraphics[width=\columnwidth]{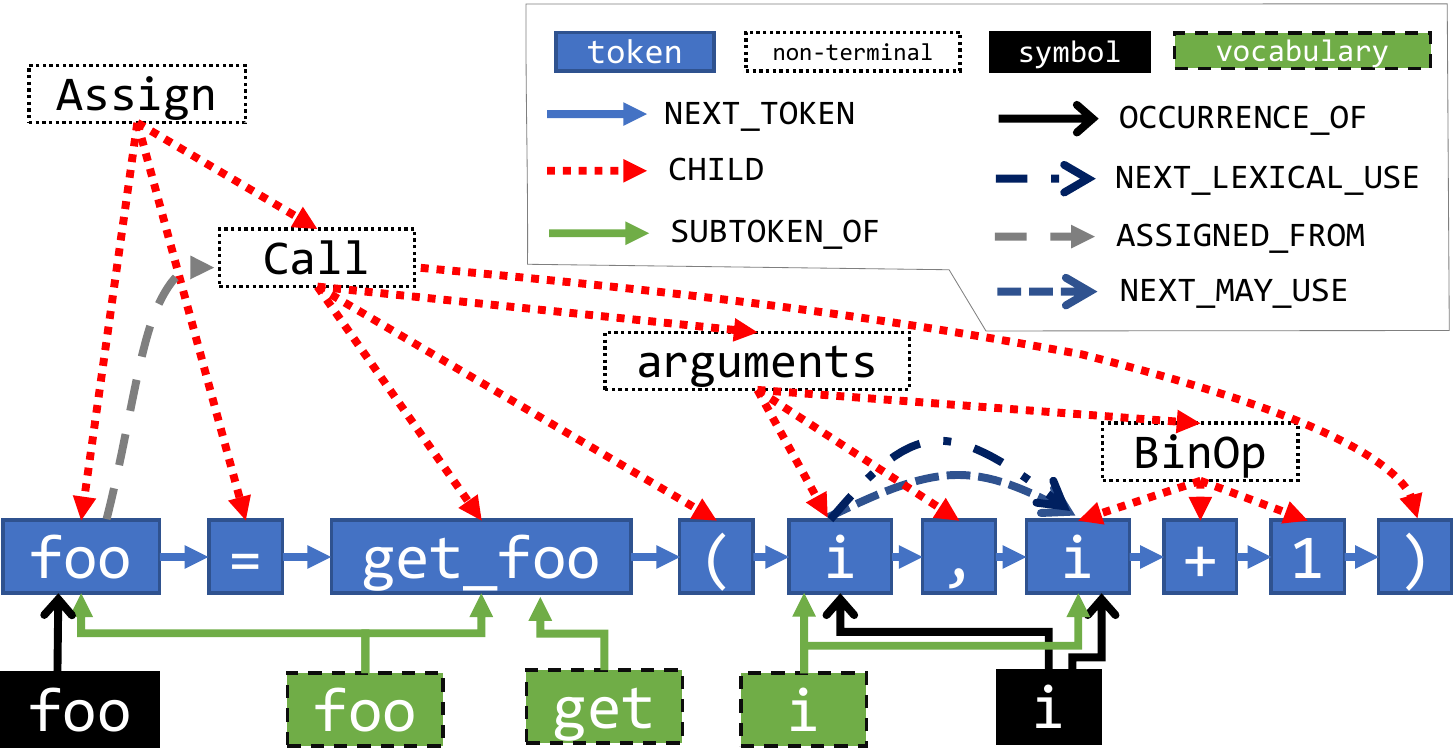}
    \caption{Sample graph for \code{foo=get\_foo(i, i+1)}
        showing different node categories and edge labels.}\label{fig:sampleGraph}\vspace{-1em}
\end{figure}

Our graphs are extracted per-file, excluding all comments, using the \href{https://github.com/python/typed_ast}{\code{typed\_ast}}
and \href{https://docs.python.org/3.7/library/symtable.html}{\code{symtable}} Python packages
by performing a dataflow analysis.
\autoref{fig:sampleGraph} illustrates such a graph.
Our graph consists of four categories of nodes:
\begin{itemize}
\item \emph{token} nodes represent the raw lexemes in the program.
\item \emph{non-terminal} nodes of the syntax tree.
\item \emph{vocabulary} nodes that represents a subtoken~\citep{cvitkovic2018open}, \ie a
word-like element which is retrieved by splitting an identifier
into parts on camelCase or pascal\_case.
\item \emph{symbol} nodes that represent a unique symbol in the symbol table,
    such as a variable or function parameter.
\end{itemize}
For a symbol $s$, we set its type embeddings to $\vect{r}_{s}=\vect{h}^{t=T}_{n_s}$
where $n_s$ is $s$'s symbol node.
Symbol nodes are similar to \citet{gilmer2017neural}'s ``supernode''.  For
functions, we introduce a symbol node for each parameter and a separate symbol
node for their return.  We combine these to retrieve its signature.

We use edges to codify relationships among nodes, which the GNN uses in the
output representations. As is usual in deep learning, we do not know the
fine-grained impact of different edges, but we do know and report
(\autoref{subsec:eval:ablation}) the impact of their ablation on the overall
performance.
\autoref{tbl:edges} details our edge labels. Though some labels appear
irrelevant from the perspective of traditional program analysis, previous work
has demonstrated that they are useful in capturing code patterns
indicative of code properties. In \autoref{tbl:edges}, we cite the work that
inspired us to use each edge label. For example, \code{NEXT\_TOKEN} is redundant
in program analysis (and would be discarded after parsing) but is quite
predictive~\citep{allamanis2018learning,hellendoorn2018deep}. Particularly
important to our approach are the \code{OCCURRENCE\_OF} edges. For example, a
variable-bound token node \code{x} will be connected to its variable symbol
node, or a member access AST node \code{self.y} will be connected to the
relevant symbol node. This edge label allows different uses of the same symbol
to exchange information in the GNN.

The AST encodes syntactic information that is traditionally used in type
inference (\eg assignment and operators) so the GNN learns about these
relationships.  Function invocations are not treated specially (\ie linked to
their definitions), since, in a partially annotated codebase, statically
resolving the receiver object and thence the function is often too imprecise.
Instead, the GNN incorporates the name of the invoked function and the names of
all its keyword arguments, which in Python take the form of
\code{foo(arg\_name=value)}.

Finally, \code{SUBTOKEN\_OF} connects all identifiers that contain a subtoken to
a unique vocabulary node representing the subtoken. As \citet{cvitkovic2018open}
found, this category of edges significantly improves a model's performance on
the task of detecting variable misuses and suggesting variable names. It
represents textual similarities among names of variables and functions, even if
these names are previously unseen. For example, a variable name \code{numNodes}
and a function name \code{getNodes}, share the same subtoken \code{nodes}. By
adding this edge, the GNN learns about the textual similarity of the different
elements of the program, capturing patterns that depend on the words used by the
developer.


\section{Quantitative Evaluation}
\label{sec:evaluation}
\newcommand{\seqTannot}{\textsc{Seq}2\textsc{Class}\xspace}
\newcommand{\seqTmetric}{\textsc{Seq}2\textsc{Space}\xspace}
\newcommand{\seqThybr}{\textsc{Seq}-\projName\xspace}
\newcommand{\pathTannot}{\textsc{Path}2\textsc{Class}\xspace}
\newcommand{\pathTmetric}{\textsc{Path}2\textsc{Space}\xspace}
\newcommand{\pathThybr}{\textsc{Path}-\projName\xspace}
\newcommand{\graphTannot}{\textsc{Graph}2\textsc{Class}\xspace}
\newcommand{\graphTmetric}{\textsc{Graph}2\textsc{Space}\xspace}
\newcommand{\graphThybr}{\projName}

\projName predicts types where traditional type inference cannot. However, some
of its predictions may be incorrect, hampering \projName' utility. In this
section, we quantitatively evaluate the types \projName predicts against two
forms of ground-truth: (a) how often the predictions match existing type
annotations (\autoref{subsec:eval:match}, \autoref{subsec:eval:ablation}) and
(b) how often the predictions pass optional type checking
(\autoref{subsec:eval:tc}).

\paragraph*{Data}
We select real-world Python projects that care about types; these are the
projects likely to adopt \projName. As a proxy, we use regular
expressions to collect 600 Python repositories from GitHub that contain at least
one type annotation. We then clone those repositories and run
\href{https://github.com/google/pytype}{pytype} to augment our corpus
with type annotations that can be inferred from a static analysis tool. To allow
pytype to infer types from imported libraries, we add to the Python environment
the top 175 most downloaded libraries\footnote{Retrieved from
\url{https://hugovk.github.io/top-pypi-packages/}. Few of
packages are removed to avoid dependency conflicts.}.

Then, we run the deduplication tool of \citet{allamanis2019adverse}.
Similar to the observations of \citet{lopes2017dejavu}, we find a substantial
number of (near) code duplicates in our corpus --- more than 133k files. We
remove all these duplicate files keeping only one exemplar per cluster of
duplicates. As discussed in \citet{allamanis2019adverse}, failing to remove
those files would significantly bias our results. We provide a Docker container
that replicates corpus construction and a list of the cloned projects (and SHAs)
at \projURL.

The collected dataset is made of 118\,440 files with a
total 5\,997\,459 symbols of which 252\,470 have a non-\code{Any}
non-\code{None} type annotation\footnote{We exclude \code{Any} and \code{None}
type annotations from our dataset.}. The annotated types are quite diverse, and
follow a heavy-tailed distribution. There are about 24.7k distinct
non-\code{Any} types, but the top 10 types are about half of the dataset.
Unsurprisingly, the most common types are \code{str}, \code{bool} and \code{int}
appearing 86k times in total. Additionally, we find only 158 types with more
than 100 type annotations, where each one of the rest 25k types are used within
an annotation less than 100 times per type, but still amount to 32\% of the
dataset. This skew in how type annotations are used illustrates the importance
of correctly predicting annotations not just for the most frequent types but for
the long tail of rarer types. The long-tail of types, consist of user-defined
types and generic types with different combinations of type arguments. Finally,
we split our data into train-validation-test set in 70-10-20 proportions.


\subsection{Quantitative Evaluation}
\label{subsec:eval:match}

Next, we look at the ability of our model to predict
ground-truth types. To achieve this, we take existing code,
erase all type annotations and aim to retrieve the original annotations.

\paragraph*{Measures}
Measuring the ability of a probabilistic system that predicts types is a
relatively new domain. For a type prediction $\tau_p$ and the ground truth type
$\tau_g$, we propose three criteria and measure the performance of a type
predicting system by computing the ratio of predictions, over all predictions,
that satisfy each criterion:
\begin{description}
    \item[Exact Match] $\tau_p$ and $\tau_g$ match exactly.
    \item[Match up to Parametric Type] Exact match when ignoring all type parameters (outermost \code{[]}).
    \item[Type Neutral] $\tau_p$ and $\tau_g$ are neutral, or interchangeable, under
    optional typing.
\end{description}
In \autoref{subsec:eval:match} and \autoref{subsec:eval:ablation}, we
approximate type neutrality. We preprocess all types seen in the corpus,
rewriting components of a parametric type whose nested level is greater than 2
to \code{Any}. For example, we rewrite \code{List[List[List[int]]]} to
\code{List[List[Any]]}. We then build a type hierarchy for the preprocessed
types. Assuming universal covariance, this type hierarchy is a lattice ordered
by subtyping $:<$. We heuristically define a prediction $\tau_p$ to be neutral
with the ground-truth $\tau_g$ if $\tau_g :< \tau_p \wedge \tau_p \neq \top$ in
the hierarchy. This approximation is unsound, but fast and scalable. Despite
being unsound, the supertype still conveys useful information, facilitating
program comprehension and searching for $\tau_g$. In \autoref{subsec:eval:tc},
we assess type neutrality by running an optional type checker. We replace
$\tau_g$ in a partially annotated program $P$ with $\tau_p$, creating a new
program $P'$, and optionally type check $P'$ to observe whether the replacement
triggers a type error. Note that an optional type checker's assessment of type
neutrality may change as $P$ becomes more fully annotated.

\paragraph*{Baselines}
The first set of baselines --- prefixed with ``\textsc{Seq}'' --- are based on
DeepTyper~\citep{hellendoorn2018deep}. Exactly as in DeepTyper, we use 2-layer
biGRUs~\citep{bahdanau2014neural} and a consistency module in between layers.
The consistency module computes a single representation for each variable by
averaging the vector representations of the tokens that are bound to the same
variable. Our models are identical to DeepTyper with the following exceptions
(a) we use subtoken-based embeddings which tend to help generalisation (b)  we
add the consistency module to the output biGRU layer, retrieving a single
representation per variable. Using this method, we compute the type embedding of
each variable.

The second set of baselines (denoted as *\textsc{Path}) are based on
code2seq~\citep{alon2018code2seq}. We adapt code2seq from its original task of
predicting sequences to predicting a single vector by using a self-weighted
average of the path encodings similar to \citet{gilmer2017neural}. For each
symbol, we sample paths that involve that tokens of that symbol and other leaf
identifier nodes. We use the hyperparameters of \citet{alon2018code2seq}.

We test three variations for the \textsc{Seq}-based, \textsc{Path}-based, and
graph-based models. The models suffixed with \textsc{Class} use the
classification-based loss (\autoref{eq:softmax classification}), those suffixed
with \textsc{Space} use similarity learning and produce a type space
(\autoref{eq:modifiedTriplet}). Finally, models suffixed with \projName use the
full loss (\autoref{eq:projLoss}). The *\textsc{Space} and *\projName models
differ only in the training objective, but are otherwise identical.

\paragraph*{Results}
\begin{table*}\centering
    \begin{tabular}{ll@{\hspace{1em}}rrr@{\hspace{1em}}rrrrr@{\hspace{1em}}} \toprule
                  &   \multirow{2}{*}{Loss}    & \multicolumn{3}{c}{\% Exact Match} &        & \multicolumn{3}{c}{\% Match up to Parametric Type} & \% Type                                                \\ \cmidrule{3-5} \cmidrule{7-9}
                  &                            & All
		  & Common & Rare                                           &
  & All        & Common & Rare & Neutral  \\ \midrule
       \seqTannot   &\autoref{eq:softmax classification}& 39.6                                & 63.8  &  4.6                                          &         & 41.2      & 64.6         & 7.6 & 30.4      \\
       \seqTmetric  &\autoref{eq:modifiedTriplet}       & 47.4                                & 62.2  & 24.5                                          &         & 51.8      & 63.7         & 32.2& 48.9      \\
       \seqThybr    &\autoref{eq:projLoss}              & 52.4                                & 71.7  & 24.9                                          &         & 59.7      & 74.2         & 39.3& 53.9      \\
       \pathTannot  &\autoref{eq:softmax classification}& 37.5                                & 60.5  &  5.2                                          &         & 39.0      & 61.1         & 7.9 & 34.0\\
       \pathTmetric &\autoref{eq:modifiedTriplet}       & 42.3                                & 61.9  & 14.5                                          &         & 47.4      & 63.6         & 24.8& 43.7\\
       \pathThybr   &\autoref{eq:projLoss}              & 43.2                                & 63.8  & 13.8                                          &         & 49.2      & 65.8         & 25.7& 44.7\\
       \graphTannot &\autoref{eq:softmax classification}& 46.1                                & 74.5  & 5.9                                           &         & 48.8      & 75.4         & 11.2& 46.9      \\
       \graphTmetric&\autoref{eq:modifiedTriplet}       & 50.5                                & 69.7  & 23.1                                          &         & 58.4      & 72.5         & 38.4& 51.9      \\
       \graphThybr  &\autoref{eq:projLoss}              & 54.6                                & 77.2  & 22.5                                          &         & 64.1      & 80.3         & 41.2& 56.3      \\
        \bottomrule\end{tabular}
    \caption{Quantitative evaluation of models measuring their ability to
    predict ground truth type annotations. Breakdown for common (seen $\ge 100$ times) and
        rare types (seen $<100$ times). Results averaged over two randomly initialised trainings.}\label{tbl:typeresults}
\end{table*}

\begin{figure*}\centering
    \begin{subfigure}[b]{0.33\textwidth}\centering
        \includegraphics[width=\columnwidth]{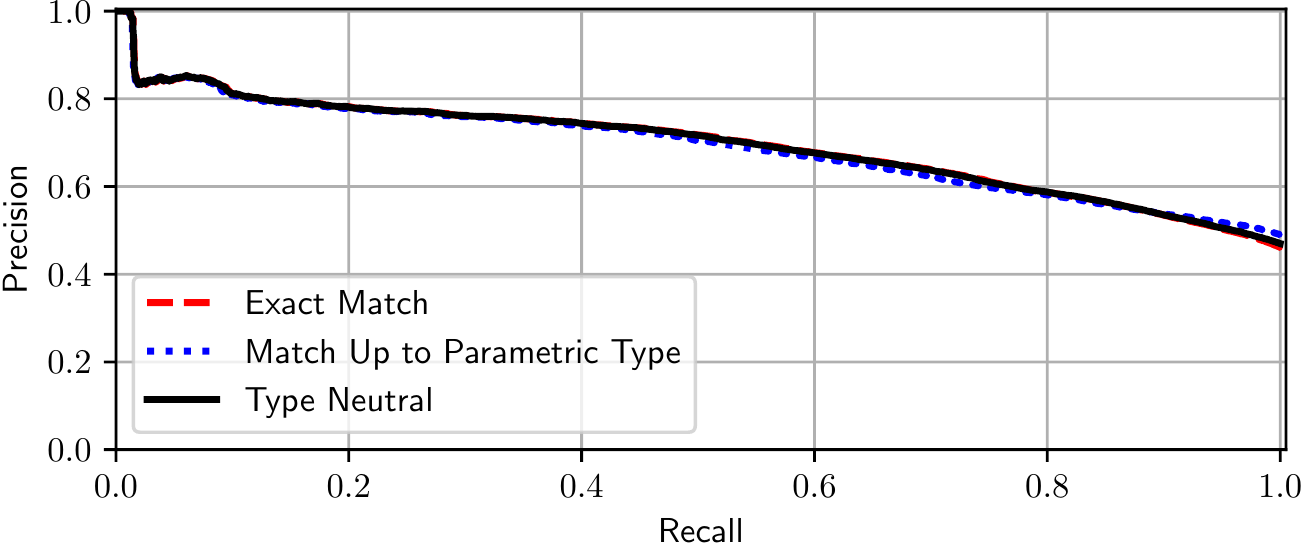}
        \caption{\graphTannot}
    \end{subfigure}
    \begin{subfigure}[b]{0.33\textwidth}\centering
        \includegraphics[width=\columnwidth]{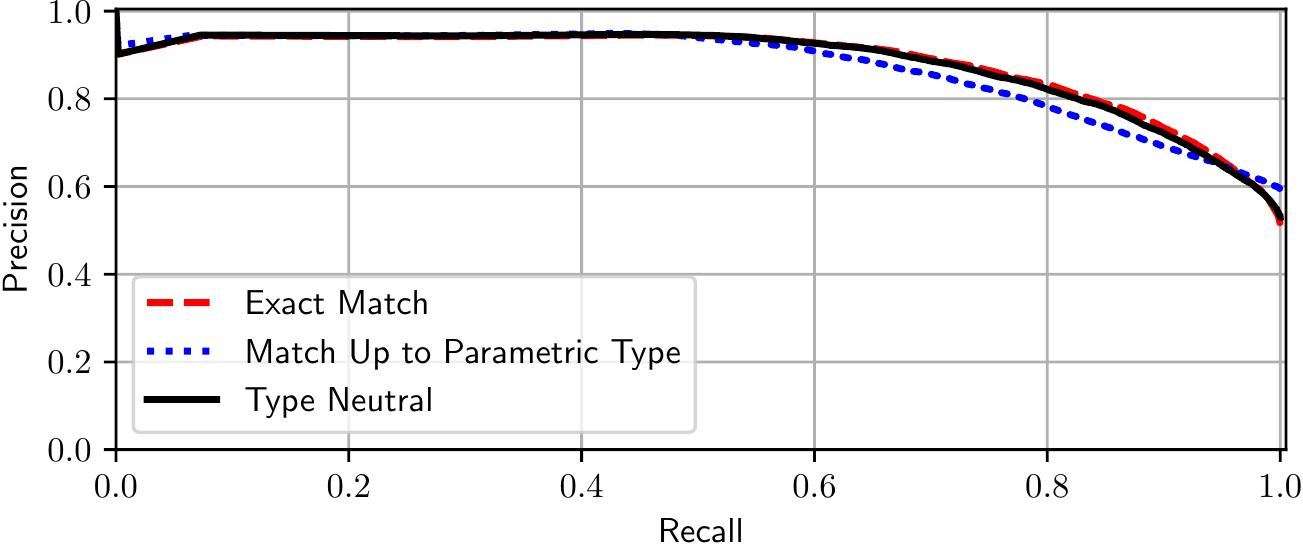}
        \caption{\graphTmetric}
    \end{subfigure}
    \begin{subfigure}[b]{0.33\textwidth}\centering
        \includegraphics[width=\columnwidth]{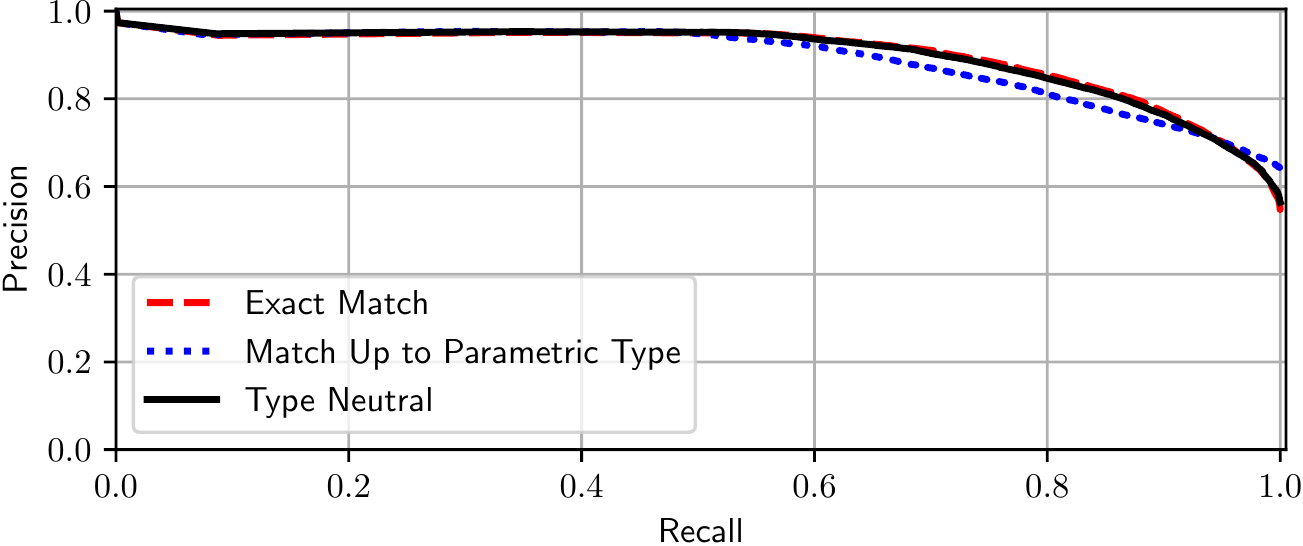}
        \caption{\graphThybr}
    \end{subfigure}
\caption{Precision-recall Curves. When filtering by confidence, \graphThybr
makes precise predictions; compared to the baselines, 95\% of the predictions
are type neutral, when \graphThybr predicts a type to 60\% of all symbols
(\ie 60\% recall). } \label{fig:prcurve}
\end{figure*}

\autoref{tbl:typeresults} shows the results of the various methods and
variations. First, it shows that the graph-based models outperform the
sequence-based and path-based models on most metrics, but not by a wide margin.
This suggests that graphs capture the structural constraints of the code
somewhat better than sequence models. When we break down the results into the
types that are often seen in our test set (we arbitrarily define types seen
fewer than 100 times to be rare), we see that the meta-learning learning methods
are significantly better at predicting the rare types and perform only slightly
worse than classification-based models on the more common types.  Combining
meta-learning and classification (\projName loss in \autoref{eq:projLoss}) yields
the best results.  The \textsc{Path}-based methods~\citep{alon2018code2seq}
perform slightly worse to the sequence-based methods. We believe that this is
because sequence models treat the problem as structured prediction
(predicting the type of multiple symbols simultaneously), whereas path-based
models make independent predictions.

\begin{figure}[tb]\centering
    \includegraphics[width=\columnwidth]{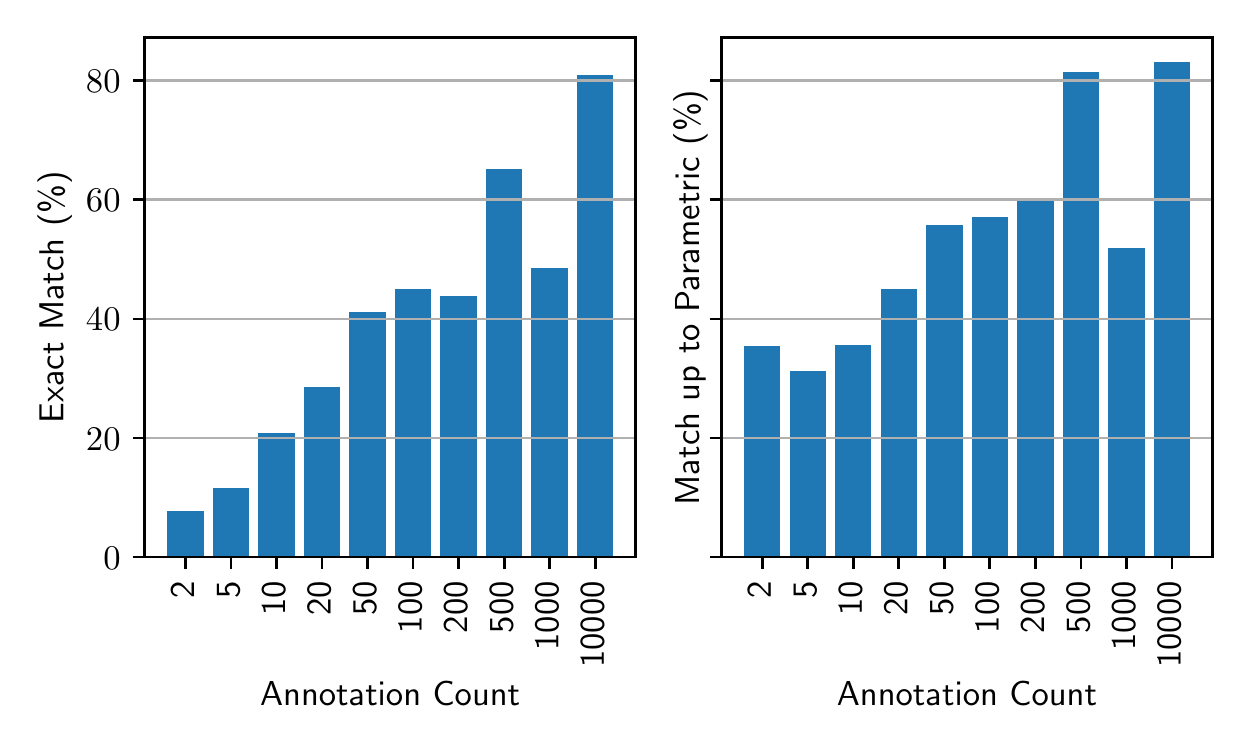}
    \caption{\graphThybr's performance bucketed by the number of
            annotations of a given type in our dataset.} \label{fig:perf by count}
\end{figure}

\autoref{fig:perf by count} breaks down the performance of \graphThybr
by the number of times each type is seen in an annotation. Although performance
drops for rare annotations the top prediction is often valid.
Since a type checker can eliminate false positives, the valid predictions 
will improve \graphThybr's performance.

The precision-recall curve of \graphThybr in \autoref{fig:prcurve} shows that it
achieves high type neutrality for high-confidence predictions compared to the
baselines. This suggests that, if we threshold on the prediction's confidence,
we can vary the precision-recall trade-off. In particular, the curve shows that
\graphThybr achieves a type neutrality of about 95\% when predicting types for
70\% of the symbols, implying that this method works well enough for eventual
integration into a useful tool. As we discuss in \autoref{subsec:eval:tc}, we
further eliminate false positives by filtering the suggestions through a type
checker, which removes ``obviously incorrect'' predictions. \autoref{tbl:per
symbol kind} shows the breakdown of the performance of \graphThybr over
different kinds of symbols. \projName seems to perform worse on variables
compared to other symbols on exact match, but not on match up to the parametric
type. We believe that this is because  in our data variable annotations are more
likely to involve generics compared to parameter or return annotations.

\begin{table}\centering
    \begin{tabular}{lrrr} \toprule
        & \multirow{2}{*}{Var} & \multicolumn{2}{c}{Func} \\ \cmidrule{3-4}
        &                           & Para & Ret\\ \midrule
\% Exact Match     & 43.5 & 53.8  & 56.9\\
\% Match up to Parametric Type & 61.8 & 57.9  & 69.5\\
\% Type Neutral & 45.5 & 55.1  & 58.9 \\ \midrule
\footnotesize Proportion of testset   &\footnotesize  9.4\% &\footnotesize 41.5\%  &\footnotesize 49.1\%\\
   \bottomrule \end{tabular}
   \caption{\projName's performance
   by the kind of symbol.}\label{tbl:per symbol kind}
\end{table}




\paragraph*{Relating Results to JavaScript} The results presented here are
numerically worse than those for JavaScript corpora presented in
JSNice~\citep{raychev2015predicting} and DeepTyper~\citep{hellendoorn2018deep}.
We posit three reasons for this difference.  First, \projName's Python dataset
and the previous work's JavaScript datasets differ significantly in the targeted
application domains.  Second, code duplicated, or shared, across the training
and test corpora of the previous work may have affected the reported
results~\citep{allamanis2019adverse}.  Third, the dynamic type systems of Python
and JavaScript are fundamentally different.  Python features a type system that
supports many type constructors and enforces strict dynamic type checking. This
has encouraged developers to define type hierarchies that exploit the error
checking it offers.  In contrast, JavaScript's dynamic type system is less
expressive and more permissive.  The detailed, and sparse, type hierarchies that
Python programs tend to have makes predicting type annotations harder in Python
than in JavaScript.

\paragraph*{Computational Speed}
The GNN-based models are significantly faster compared to RNN-based models. On a
Nvidia K80 GPU, a single training epoch takes 86sec for the GNN model, whereas
it takes 5\,255sec for the biRNN model. Similarly, during inference the GNN
model is about 29 times faster taking only 7.3sec per epoch (\ie less than 1ms
per graph on average). This is due to the fact that the biRNN-based models
cannot parallelise the computation across the length of the sequence and thus
computation time is proportional to the length of the sequence, whereas GNNs
parallelise the computation but at the cost of using information that is only a
few hops away in the graph. However, since the graph (by construction) records
long-range information explicitly (\eg data flow) this does not impact the
quality of predictions.

\paragraph*{Transformers}
An alternative to RNN-based models are
transformers~\citep{vaswani2017attention}, which have recently shown exceptional
performance in natural language processing. Although transformers can be
parallelised efficiently, their memory requirements are quadratic to the
sequence length. This is prohibitive for even moderate
Python files. We test small transformers (replacing the biGRU of our DeepTyper)
and remove sequences with more than 5k tokens, using a mini-batch size of 2.
The results did \emph{not} improve on DeepTyper. This may be because
transformers often require substantially larger quantities of data to outperform other models.

\subsection{Ablation Analysis}
\label{subsec:eval:ablation}
Now, we test \projName's performance when varying different elements of its
architecture. Our goal is \emph{not} to be exhaustive, but to illustrate
how different aspects affect type prediction.
\autoref{tbl:edge ablation} shows the results of the ablation study, where we
remove some edge labels from the graph at a time and re-train our neural network
from scratch. The results illustrate the (un)importance of each aspect of the
graph construction, which \autoref{sec:implementation} detailed. First, if we
simply use the names of the symbol nodes the performance drops significantly and
the model achieves an exact match of 37.6\%. Nevertheless, this is a significant
percentage and attests to the importance of the noisy, but useful, information
that identifiers contain. Removing the syntactic edges, \code{NEXT\_TOKEN} and
\code{CHILD}, also reduces the model's performance, showing that our model can
find patterns within these edges. Interestingly, if we just remove
\code{NEXT\_TOKEN} edges, we still see a performance reduction, indicating that
tokens --- traditionally discarded in formal analyses, since they are redundant
--- can be exploited to facilitate type prediction. Finally, the data-flow
related edges, \code{NEXT\_LEXICAL\_USE} and \code{NEXT\_MAY\_USE}, have
negligible impact on the overall performance. The reason is that for type
prediction, the order of a symbol's different uses does not matter. Therefore,
representing these uses in a sequence (\code{NEXT\_LEXICAL\_USE}) or a tree
(\code{NEXT\_MAY\_USE}) offers no additional benefits. Simply put,
\code{OCCURRENCE\_OF} subsumes \code{NEXT\_*USE} in our problem setting.

\begin{figure}[tb]
    \includegraphics[width=\columnwidth]{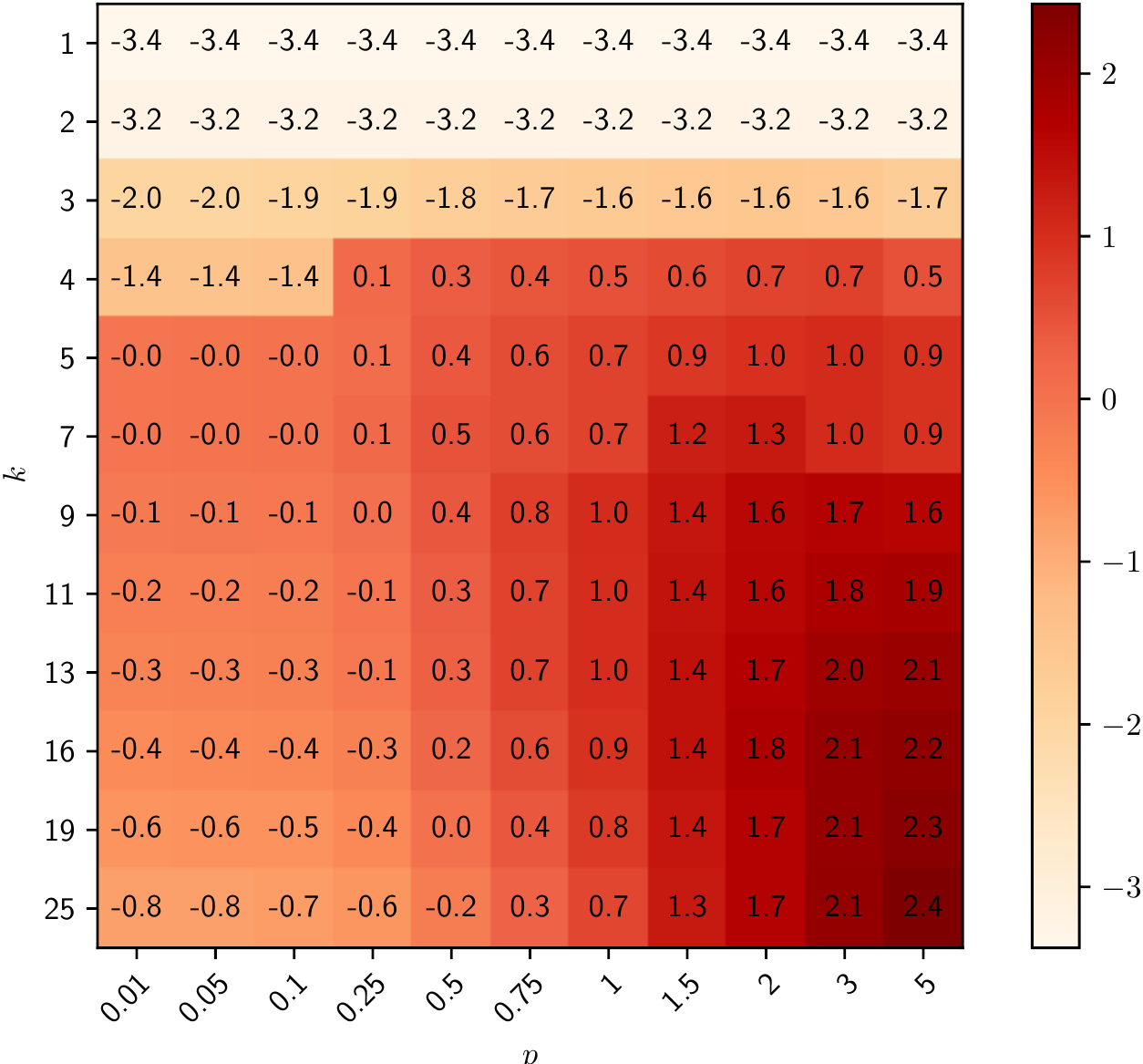}
    \caption{Absolute difference in match up to parametric type for \projName w.r.t. median
    for various $k$ and $p$ in \cref{eq:knn prob}.}\label{fig:knn ablation}
\end{figure}

\autoref{tbl:edge ablation} also shows how the performance of \projName varies
with different token representations for the initial node states of the GNN. We
test two variations: Token-level representations, where each lexeme gets a
single embedding as in \citet{hellendoorn2018deep}, and character-level
representations that use a 1D convolutional neural
network~\citep{kim2016character} to compute a node representation from its
characters. The results
suggest that the initial representation does \emph{not}
have a big difference, with \projName's subtoken-level models having a small
advantage, whereas the character-level CNN models performing the worst, but
with a small difference.

Finally, \autoref{fig:knn ablation} shows \graphThybr's performance when varying
$k$ and $p$ of \autoref{eq:knn prob}. The results
suggest that larger values for $k$ give better results and a larger $p$ also
helps, suggesting that looking at a wider neighbourhood in the type map but
accounting for distance can yield more accurate results.

\begin{table}[tb]\centering
    \begin{tabular}{lrr} \toprule
        Ablation            & Exact Match& Type Neutral  \\ \midrule
    Only Names (No GNN)     & 38.8\%   &  40.4\% \\  
    No Syntactic Edges      & 53.7\%   &  55.6\% \\  
    No \code{NEXT\_TOKEN}   & 54.7\%   &  56.3\% \\  
    No \code{CHILD}         & 48.4\%   &  50.2\% \\  
    No \code{NEXT\_*USE}    & 54.7\%   &  56.4\%  \\  
    Full Model -- Tokens    & 53.7\%   &  55.4\%\\  
    Full Model -- Character & 53.4\%   &  55.0\%\\ 
    Full Model -- Subtokens & 54.6\%   &  56.3\%  \\ 
    \bottomrule
    \end{tabular}
    \caption{Ablations of \graphThybr when removing
        edges from the graph or varying
        the initial node representation.}\label{tbl:edge ablation}
\end{table}

\subsection{Correctness Modulo Type Checker}
\label{subsec:eval:tc}

So far, we treated existing type annotations --- those manually added by the
developers or those inferred by pytype --- as the ground-truth. However, as we
discuss in \autoref{sec:qual:eval}, some annotations can be wrong, \eg because
developers may not be invoking a type checker and many symbols are \emph{not}
annotated. To thoroughly evaluate \projName, we now switch to a different ground
truth: optional type checkers. Though optional type checkers reason over only a
partial context with respect to a fully-typed program and are generally unsound,
their best-effort is reasonably effective in practice~\citep{gao2017type}. Thus,
we take their output as the ground-truth here.

Specifically, we test one type prediction at a time and pick the top prediction
for each symbol. For each prediction $\tau$ for a symbol $s$ in an annotated
program $P$, we add $\tau$ to $P$ if $s$ is not annotated, or replace the
existing annotation for $s$ with $\tau$, retaining all other annotations in $P$.
Then, we run the optional type checker and check if $\tau$ causes a type error.
We repeat this process for all the top predictions and aggregate the results.
This experiment reflects the ultimate goal of \projName: helping developers
gradually move an unannotated or partially annotated program to a fully
annotated program by adding a type prediction at a time.

We consider two optional type checkers: \href{http://mypy-lang.org/}{mypy} and
\href{https://github.com/google/pytype}{pytype}. In 2012, mypy introduced
optional typing for Python and strongly inspires Python's annotation syntax.
Among other tools, pytype stands out
because it employs more powerful type inference and more closely reflects
Python's semantics, \ie it is less strict in type checking than a
traditional type checker, like mypy. Both type checkers are popular and actively
maintained, but differ in design mindset, so we include both to cover different
philosophies on optional typing.

To determine how often \projName's type predictions are correct, we first
discard any programs which fail to type check \emph{before} using \projName,
since they will also fail even when \projName's type predictions are correct.
Since mypy and pytype also perform other static analyses, such as linting and
scope analysis, we need to isolate the type-related errors. To achieve this, we
comb through all error classes of mypy and pytype and, based on
their description and from first principles, decide which errors relate
to types. We then use these type-related error classes to filter the programs in
our corpus. This filter is imperfect:  some error classes, like
``\href{https://mypy.readthedocs.io/en/stable/error_code_list.html#miscellaneous-checks-misc}{\code{[misc]}}''
in mypy, mix type errors with other errors. To resolve this, we sample the
filtered programs and manually determine whether the sampled programs have type
errors. This process removes 229 programs for mypy and 10 programs for pytype
that escaped the automated filtering based on error classes. We provide
more information at \projURL. After preprocessing the
corpus, we run mypy and pytype on the remaining programs, testing one prediction
at a time. We skip type predictions which are \code{Any}, or on which mypy
or pytype crashes or spends more than 20 seconds. In total, we assess 350,374
type predictions using mypy and 85,732 using pytype.

\begin{table}[t]
    \centering
    \begin{tabular}{lll r r r r r}
        \toprule
        \multicolumn{2}{c}{Annotation}&&  \multicolumn{2}{c}{mypy} &  & \multicolumn{2}{c}{pytype}           \\ \cmidrule{1-2} \cmidrule{4-5} \cmidrule{7-8}
        Original  & Predicted               &&  Prop.    & Acc.           &  & Prop.    & Acc. \\
        \midrule
        $\epsilon$ & $\tau$ &&  \footnotesize 95\%   &  89\%         &  & \footnotesize 94\%   & 83\% \\
        $\tau$ & $\tau'$    &&  \footnotesize 3\%    &  85\%         &  & \footnotesize 3\%   & 63\% \\
        $\tau$ & $\tau$     &&  \footnotesize 2\%    & 100\%         &  & \footnotesize 3\%   & 100\% \\ \midrule
        \multicolumn{2}{c}{Overall} &&  \footnotesize 100\%  &  89\%         &  & \footnotesize 100\%  & 83\% \\
        \bottomrule
    \end{tabular}
    \caption{Type checking accuracy of \projName modulo mypy and pytype.
     A prediction is incorrect if it causes a type error.
     Mypy and pytype experience timeouts on different programs, hence the discrepancy between the proportion of each case.}
    \label{tbl::tc:results}
\end{table}

\begin{figure}[t]\centering
    \includegraphics[width=\columnwidth]{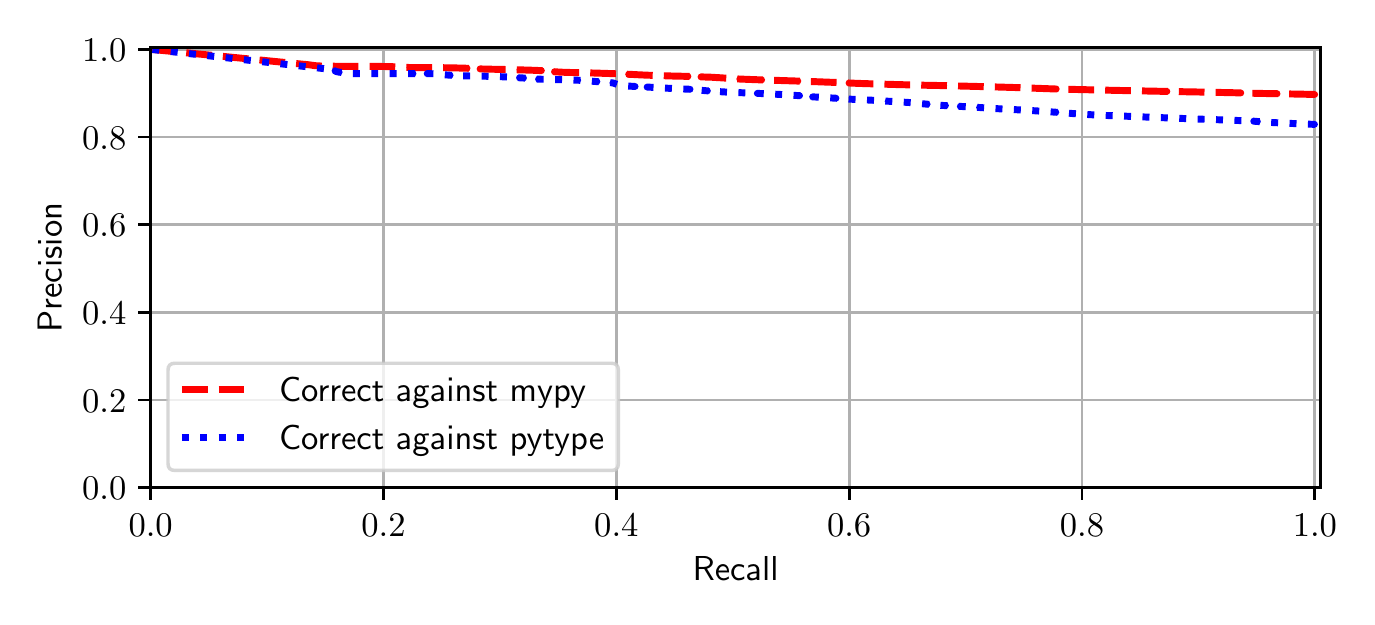}
    \caption{Precision-recall curve for the type checking experiment.
     We deem \projName unable to suggest a type if the probability of a type prediction is below a threshold.}
    \label{fig:tcprcurve}
\end{figure}

\autoref{tbl::tc:results}
presents the results of applying mypy and pytype to the top type predictions. In
general, 89\% and 83\% of \projName's predictions do not cause a type error in
mypy and pytype, respectively. This demonstrates that the type predictions are
commonly correct with respect to optional typing. We then group the predictions
into three categories: $\epsilon \rightarrow \tau$ where \projName suggests a
type to a previously unannotated symbol, $\tau \rightarrow \tau'$ where it
suggests a type that is different from the original annotation, and $\tau
\rightarrow \tau$ where the suggested type is identical with the original
annotation. As the $\epsilon \rightarrow \tau$ row illustrates, most of the
symbols, whose types \projName is able to predict, are untyped even after
pytype's type inference. This indicates that \projName has a wide application
domain.

For mypy, 3\% of \projName's predictions differ from the original annotations.
Though different, some of these predictions might actually be correct
(\autoref{sec:qual:eval}). Further analysis reveals that 33\% of these
predictions are a supertype of the original one (less precise but
interchangeable) and 2\% are more specific (and potentially but not certainly
incorrect). Mypy produces a type error for 22\% of them, which shows that
optional type checkers can effectively improve the quality
\projName's suggestion, by filtering false positives. The $\tau \rightarrow
\tau$ case is a sanity check: the input programs do not have type errors, by
construction of this experiment, so when \projName predicts
the same annotations, they pass type checking.

Finally, \autoref{fig:tcprcurve} investigates \projName' precision and recall.
By varying the confidence threshold on the predictions, we can trade precision
for recall. \projName maintains a good trade-off between precision and recall.
For example, when it predicts a type for 80\% of all symbols, 90\% of the
predictions are correct with respect to mypy.

\section{Qualitative Evaluation}
\label{sec:qual:eval}

To better understand the performance of \projName and the expressivity
of the types it can infer, we manually analyse its
predictions, before a type checker filters them. Our goal is \emph{not}
to be exhaustive but to convey cases that indicate opportunities for future
research. 


We begin by exploring how complex a type expression \projName can reliably
infer. By complex, we mean a deeply nested parametric type such as
\code{Set[Tuple[bool}, \code{Tuple[UDT}, \code{...]]]}, where UDT denotes a
user-defined type.  In principle, \projName can learn to predict these types.
However, such types are extremely rare in our dataset: about 30\% of our
annotations are parametric and, of them, 80\% have depth one and 19\% have depth
two, with more deeply nested types mostly appearing once.  For our evaluation,
we built the type map over the training and the validation sets.  Since these
complex types appear once (and only in our test set), they do not appear in the
type map and \projName cannot predict them.  We believe that deeply nested
parametric types are so rare because developers prefer to define and annotate
UDTs rather than deeply nested types which are
hard-to-understand.  Unfortunately, \projName finds UDTs hard to predict.
Improved performance on the task will require machine learning methods that
better understand the structure of UDTs and the semantics of their names.

We now look at the most confident errors, \ie cases where \projName
confidently predicts a non-type neutral type.
\projName
commonly confuses variables with collections whose elements have the same type
as the variables, confusing \code{T} and \code{Optional[T]}, for various
concrete \code{T}, and predicts the wrong
type\footnote{\href{https://docs.python.org/3/library/typing.html\#typing.Optional}{\code{Optional[T]}}
conveys a nullable variable, \ie \code{Union[T, None]}.}.  Similarly, when the
ground truth type is a \code{Union}, \projName often predicts a subset of the
types in the union. This suggests that the type space is not learning to
represent union types. For example, in
\href{https://github.com/rembo10/headphones/blob/5283b48736cda701416aad11467649224c6e7080/lib/beets/mediafile.py#L1208}{rembo10/headphones},
\projName predicts \code{Optional[int]} where the ground truth is
\code{Optional[Union[float, int, str]]}.  Adding intraprocedural relationships
to \graphThybr, especially among different code files, may help resolve such
issues.

We also identified a few cases where the human type annotation is wrong.  For
example, in \href{https://github.com/pytorch/fairseq}{\code{PyTorch/fairseq}}, a
sequence-to-sequence modelling toolkit that attracts more than 7.3k stars on
GitHub, we found three parameters representing tensor dimensions annotated as
\code{float}. \projName, having seen similar code and similarly named variables,
predicts with 99.8\% confidence that these parameters should be annotated as
\code{int}. We submitted two pull requests covering such cases:
one\footnote{\url{https://github.com/pytorch/fairseq/pull/1268}} to
\href{https://github.com/pytorch/fairseq}{\code{PyTorch/fairseq}} and
one\footnote{\url{https://github.com/allenai/allennlp/pull/3376}} to
\href{https://github.com/allenai/allennlp}{\code{allenai/allennlp}}, a natural
language processing library with more than 8.2k stars. Both have been merged.
\emph{Why did the type checker fail to catch these errors?} The
problem lies with the nature of optional typing.  It can only reason locally
about type correctness;  it only reports an error if it finds a local type
inconsistency. When a function invokes an unannotated API, an optional type
checker can disprove very few type assignments involving that call. This is an
important use-case of \projName, where, due to the sparsely annotated nature of
Python code, incorrect annotations can go undetected by type checkers.

In some cases, \projName predicts a correct, but more specific type, than
the original type annotation or the one inferred by pytype. For example,
in an
\href{https://github.com/ansible/ansible/blob/94c23136be35ca5334945a66c1342863dd026fa4/lib/ansible/modules/network/fortios/fortios_wireless_controller_vap.py#L1110}{Ansible
function}, pytype inferred \code{dict}, whereas \projName predicted
the more precise \code{Dict[str, Any]}. We believe that this problem arises due
to pytype's design, such as preferring conservative approximations,
which allow it to be a practical type inference engine.

A third source of disagreement with the ground truth is confusing \code{str} and
\code{bytes}. These two types are conceptually related: \code{bytes} are raw
while \code{str} is ``cooked'' Unicode. Indeed, one can always encode a
\code{str} into its \code{bytes} and decode it back to a \code{str}, given some
encoding such as ASCII or UTF-8.  Python developers also commonly confuse
them~\citep{stackoverflowPythonEncode}, so it is not surprising that \projName
does too.  We believe that this confusion is due to the fact that variables and
parameters of the two types have a very similar interface and they usually share
names.  Resolving this will require a wider understanding how the underlying
object is used, perhaps via an approximate intraprocedural analysis.


Finally, \projName confuses some user-defined types. Given the sparsity of these
types, this is unsurprising. Often, the confusion is in conceptually related
types. For example, in
\href{https://github.com/awslabs/sockeye/blob/c78db6e651c6589e8a01aef03f91c68abb91b7b5/test/unit/test_inference.py#L504}{awslabs/sockeye},
a variable annotated with \code{mx.nd.NDArray} is predicted to
be \code{torch.Tensor}. Interestingly, these two types represent tensors but in
different machine learning frameworks
(\href{https://mxnet.incubator.apache.org/}{MxNet} and
\href{https://pytorch.org/}{PyTorch}).

\section{Related Work}

Recently, researchers have been exploring the
application of machine learning to code~\citep{allamanis2018survey}. This
stream of work focuses on capturing (fuzzy) patterns in code to perform
tasks that would not be possible with using only formal methods. Code
completion~\citep{raychev2014code,hindle2012naturalness,hellendoorn2017deep,karampatsis2019maybe}
is one of the most widely explored topics. Machine learning models of code are
also used to predict names of variables and
functions~\citep{allamanis2014learning,allamanis2015suggesting,alon2018code2vec,alon2018code2seq,bavishi2017context2name},
with applications to
deobfuscation~\citep{raychev2015predicting,vasilescu2017recovering} and
decompilation~\citep{he2018debin,david2019neural,lacomis2019dire}. Significant
effort has been made towards automatically generating documentation from code or
vice
versa~\citep{iyer2016summarizing,barone2017parallel,alon2018code2seq,fernandes2019structured}.
These works demonstrate that machine learning can effectively capture
patterns in code.

\paragraph*{Program Analysis and Machine Learning}
Within the broader area of program analysis, machine learning has been utilised
in a variety of settings. Recent work has focused on detecting specific kinds of
bugs, such as variable misuses and argument
swappings~\citep{allamanis2018learning,cvitkovic2018open,pradel2017deep,vasic2019neural,rice2017detecting}
showing that ``soft'' patterns contain valuable information that can catch many
real-life bugs that would not be otherwise easy to catch with formal program
analysis. Work also exists in a variety of program analysis domains.
\citet{si2018learning} learn to predict loop invariants, \citet{mangal2015user}
and \citet{heo2019continuously} use machine learning methods to filter false
positives from program analyses by learning from user feedback.
\citet{chibotaru2019scalable} use weakly supervised learning to learn taint
specifications and \citet{defreez2018path} mine API call sequence
specifications.

\paragraph*{Code Representation in Machine Learning}
Representing code for consumption in machine learning is a central research
problem. Initial work viewed code as a sequence of
tokens~\citep{hindle2012naturalness,hellendoorn2017deep}. The simplicity of this
representation has allowed great progress, but it misses the opportunity to
learn from code's known and formal structure. Other work based their
representation on
ASTs~\citep{maddison2014structured,bielik2016phog,allamanis2015bimodal,alon2018code2vec,alon2018code2seq}
--- a crucial step towards exploitation of code's structure. Finally, graphs,
which \projName employs, encode a variety of complex relationships among the
elements of code. Examples of include the early works of
\citet{kremenek2007factor} and \citet{raychev2015predicting} and variations of
GNNs~\citep{allamanis2018learning,cvitkovic2018open,heo2019continuously}.

\paragraph*{Optional Typing}
At the core of optional typing~\citep{bracha2004pluggable} lie
optional type annotations and pluggable type checking, combining
static and dynamic typing, but without providing soundness guarantees.
Therefore, it is often called ``unsound gradual
typing''~\citep{takikawa2016sound, muehlboeck2017sound}. Optional typing
has not been formalised and its realisation varies across languages and type
checkers. For example, TypeScript checks and then erases type annotations,
whereas Python's type annotations simply
decorate the code, and are used by external type checkers.

\section{Conclusion}
We presented a machine learning method for predicting types in dynamically
typed languages with optional annotations and realised it for
Python.  Type inference for many dynamic languages, like Python, must often
resort to the \code{Any} type.  Machine learning methods, like the one
presented here, that learn from the rich patterns within the code, including
identifiers and coding idioms, can provide an approximate, high-precision and
useful alternative.


\begin{acks}
We thank the reviewers for their useful feedback, and  M. Brockschmidt, J.V. Franco for
useful discussions. This work was partially supported by EPSRC grant EP/J017515/1.
\end{acks}

\bibliographystyle{ACM-Reference-Format}
\balance
\bibliography{bibliography/bibliography}

\end{document}